\documentclass[10pt,conference]{IEEEtran}
\IEEEoverridecommandlockouts
\usepackage{cite}
\usepackage{amsmath,amssymb,amsfonts}
\usepackage{algorithmic}
\usepackage{graphicx}
\usepackage{textcomp}
\usepackage{colortbl}
\usepackage{multirow}
\usepackage{xcolor}
\def\BibTeX{{\rm B\kern-.05em{\sc i\kern-.025em b}\kern-.08em
    T\kern-.1667em\lower.7ex\hbox{E}\kern-.125emX}}

\usepackage{svg}
\usepackage{listings}
\usepackage{float}
\usepackage{amsmath}
\lstdefinelanguage{pseudo}
{
    morekeywords={func,for,in,return,is,not,if,else, struct},
    morecomment=[l]{\#},
}

\newcommand{\ourFuzzer}{\emph{PrescientFuzz}}

\begin{document}

\title{PrescientFuzz: A more effective exploration approach for grey-box fuzzing}

\author{\IEEEauthorblockN{Daniel Blackwell}
\IEEEauthorblockA{\textit{Department of Computer Science} \\
\textit{University College London}\\
London, UK \\
daniel.blackwell.14@ucl.ac.uk}
\and
\IEEEauthorblockN{David Clark}
\IEEEauthorblockA{\textit{Department of Computer Science} \\
\textit{University College London}\\
London, UK \\
david.clark@ucl.ac.uk}
}

\maketitle

\begin{abstract}
Since the advent of AFL, the use of mutational, feedback directed, grey-box fuzzers has become critical in the automated detection of security vulnerabilities. 
A great deal of research currently goes into their optimisation, including improving the rate at which they achieve branch coverage early in a campaign. 
We produce an augmented version of LibAFL's `fuzzbench' fuzzer, called PrescientFuzz, that makes use of semantic information from the target program's control flow graph (CFG). 
We develop an input corpus scheduler that prioritises the selection of inputs for mutation based on the proximity of their execution path to uncovered edges. 
Simple as this idea is, PrescientFuzz leads all fuzzers using the Google FuzzBench at the time of writing -- in both average code coverage and average ranking, across the benchmark SUTs. 
Whilst the existence of uncovered edges in the CFG does not guarantee their feasibility, the improvement in coverage over the state-of-the-art fuzzers suggests that this is not an issue in practice.
\end{abstract}

\section{Introduction}

Fuzzing is an effective testing technique for finding bugs and vulnerabilities in software.
Grey-box fuzzing leverages program coverage to more effectively explore program functionality than the early black-box fuzzing approaches.
Nonetheless, even the most effective fuzzer cannot find bugs in code that its generated inputs never cover.
In this chapter, we propose an additional layer of fuzzer feedback, making use of knowledge of the target program's semantics -- specifically, the structure of its control flow graph (CFG) -- to improve the rate of coverage discovery.

We detail the design of an input scheduler. This selects inputs for mutation based on heuristics about the number of uncovered basic blocks that border its execution path; and we then expand this idea to include the complete set of all (statically) reachable uncovered blocks within the CFG.
Our approach provides the fuzzer itself with full knowledge of the CFG. 
The metrics used by the scheduler are recomputed repeatedly at runtime, meaning that it adapts dynamically as new coverage is achieved during the fuzzing campaign.
This is somewhat similar to recent \emph{directed} grey-box fuzzing approaches. Unlike that line of research we do not employ a set of fixed targets, but merely seek to optimise coverage as rapidly as possible.
The dynamic approach also means that we are able to deal with indirect branching, as branch targets for each individual execution can be determined by examining the coverage feedback; this is vital for object-oriented software that makes significant use of dynamic dispatch.

We create an implementation of the described scheduler and build it into a fuzzer, \ourFuzzer{}, which we evaluate on the FuzzBench benchmark suite \cite{metzman2021fuzzbench}; here it achieves more program coverage across the aggregated set of benchmarks than any other publicly listed fuzzer.

The feedback mechanism is implemented using an LLVM compiler pass, so is immediately available to target programs written in languages with an LLVM front-end, such as C, C++, Julia, Kotlin and Rust.
The scheduler as described is generic enough that it could be implemented for any grey-box fuzzer which should see immediate benefits.

Finally, we believe that our feedback approach has additional benefits.
For example it can be used to improve the heuristics for determining when to end a fuzzing campaign.
When doing concolic fuzzing, it could also help determine the best conditionals to solve in order to maximise exploration potential.

\section{Generalisable Techniques in Fuzzing}

\begin{figure}[t]
    \centering
    \includegraphics[width=\linewidth]{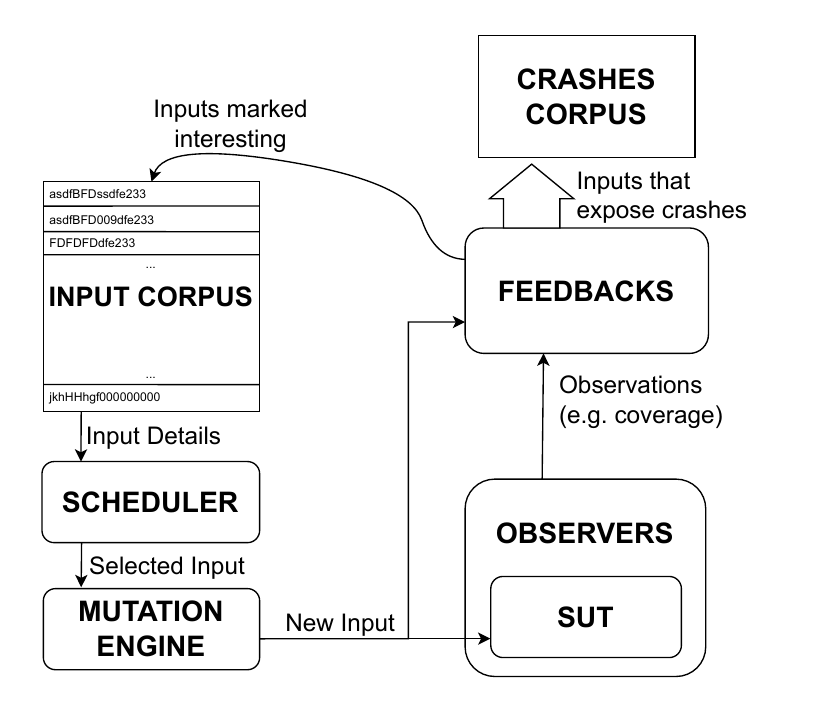}
    \caption{Typical architecture of a LibAFL-based fuzzer}
    \label{fig:fuzzer_arch}
\end{figure}

In this section, we introduce a common architecture for modern fuzzers, and discuss a number of techniques that have proven useful enough to be adopted into the current state-of-the-art grey-box fuzzers.
Like our approach, these techniques are abstract enough to be applicable to many fuzzers, rather than being a particular implementation specific optimisation.

Figure \ref{fig:fuzzer_arch} shows the architecture of a LibAFL-based fuzzer.
Note that LibAFL offers the ability to build black-, grey- and white-box fuzzers; the diagram omits the symbolic execution runtime and solver required for white-box fuzzing.
There are a number of different avenues to be explored when it comes to improving the performance of a fuzzer; the input scheduler, mutation engine, observers and feedback.
For example, the difference between black-box and grey-box fuzzers is the addition of a coverage observer (and feedback).
The following paragraphs list some well-known approaches that have been applied successfully in the past, with an aim to demonstrate how the different architectural components can be modified.
Unlike all of these, our approach combines static knowledge of the SUT's semantics (the control flow graph of the SUT), with dynamic coverage feedback, to create a new advantage.

AFLFast \cite{bohme2016coverage} extended AFL with `power schedules', in essence moving AFL from its original round-robin \emph{scheduler} to one of several specified in the paper.
Of the six proposed schedules, ``Fast'' is the most commonly used and has been re-implemented in both mainline AFL++ and LibAFL. 

Redqueen \cite{aschermann2019redqueen} is the name of a grey-box fuzzer that aims to help pass difficult conditional checks. 
It does so by searching for `input to state correspondence'; that is, relationships between parts of the input and the conditional check values.
To do this, it provides an additional \emph{observer} that records the values of any non-constant comparands at conditional checks, and the \emph{feedback} is passed on to the \emph{mutation engine}, which can try inserting the values into the input.
For example, when fuzzing a png decoder without a valid png file as seed, the first obstacle is the 8 byte magic number expected at the very start of the file.
Redqueen is able to detect that these 8 bytes from the input are used in a comparison, and tries replacing these with the value of the other comparand. 
Versions of this approach are also implemented in mainline AFL++ and LibAFL.

MOpt \cite{lyu2019mopt} is a technique that improves the \emph{mutation engine} by optimising the schedule for applying mutation operators, prioritising those that have been working well so far in the campaign.
This can be very effective for certain types of program.
The commonly used set of mutators include random bitflips, inserting constant values and splicing.
For something like a programming language parser, random bitflips are likely to be less useful than splicing; and MOpt can capitalise on this.
Again, this technique has been re-implemented in mainline AFL++ and LibAFL.


Directed grey-box fuzzing is an idea introduced by AFLGo \cite{bohme2017directed}; here a set of code points are marked as targets, and the fuzzer aims to reach these.
This is particularly useful for reproducing bugs when given access only to a stack trace, and testing patches.
In order to do so, distances are computed at compile time indicating the least number of edges between a given basic block and any targets -- inputs that get closer to the target are prioritised for mutation.
The distance acts as a form of \emph{feedback}, and the \emph{scheduler} is adapted to prioritise inputs based on this. 
Our approach bears similarities to directed grey-box fuzzing in that it uses distance metrics computed from the SUT's (System Under Test) control flow graph, however it differs in that it has no specific target at which it is being `directed'.
Instead, it aims to explore as much of the SUT's functionality as possible by maximising total coverage. 

Sanitizers introduce additional \emph{feedback} by extending the error detection oracle that the fuzzer has.
To do this they build some sanity checks into the SUT that causes it to crash if any are violated.
For example, AddressSanitizer \cite{serebryany2012addresssanitizer} can find memory errors such as use-after-free, buffer overflows and memory leaks; and MemorySanitizer can find uninitialised memory reads.
These sanitizers are implemented by compiling additional checks into the SUT.

Fishfuzz \cite{zheng2023fishfuzz} is a directed grey-box fuzzer that uses sanitizer checks as goals; this allows it to aim for code areas that are more likely to trigger a crash in the SUT.
It has proven to be an effective approach for discovering bugs.

\section{Our Approach}

Here we describe the approach used by \ourFuzzer{} for scheduling inputs for selection from the corpus.
Note that the specific details of our implementation are discussed separately in Section \ref{sec:impl}.

To quote the Fuzzbench paper \cite{metzman2021fuzzbench}, which introduces a comprehensive benchmark set for evaluating fuzzers: \emph{A fuzzer can only detect a bug if first it manages to cover the code where the bug is located.}
Thus we propose a technique to improve the rate at which coverage is discovered; particularly at the early stages of fuzzing campaigns.

To illustrate the utility of our approach we now present a worked example. 
Suppose we have a function that we wish to fuzz as follows:

\begin{figure}[H]
    \begin{lstlisting}[language=C]
int example(int in1, int in2, int in3) {
    int res = 0;
    if (in1 > 15) {
        res = 1;
    } else if (in1 < 2) {
        res = 2;
    } else if (in1 < 4) {
        res = 3;
    } else if (in1 < 8) {
        res = 4;
    } else { // 8 < i <= 15
        res = 5;
    // Hard check to break through
    if (in1 ^ in2 == 0xDEADBEEF) {
        switch (in3) {
            case 0: res = 6; break;
            case 1: res = 7; break;
            case 2: res = 8; break;
            case 3: abort(); break;  // crash the program!
            default: res = 9; break;
        }
    }
    
    return res;
}
    \end{lstlisting}
    \caption{An example program in the C programming language}
    \label{fig:sourcecode}
\end{figure}

The control flow graph for this particular example is shown in figure \ref{fig:cfg_empty}:

\begin{figure}[H]
    \includegraphics[width=\linewidth]{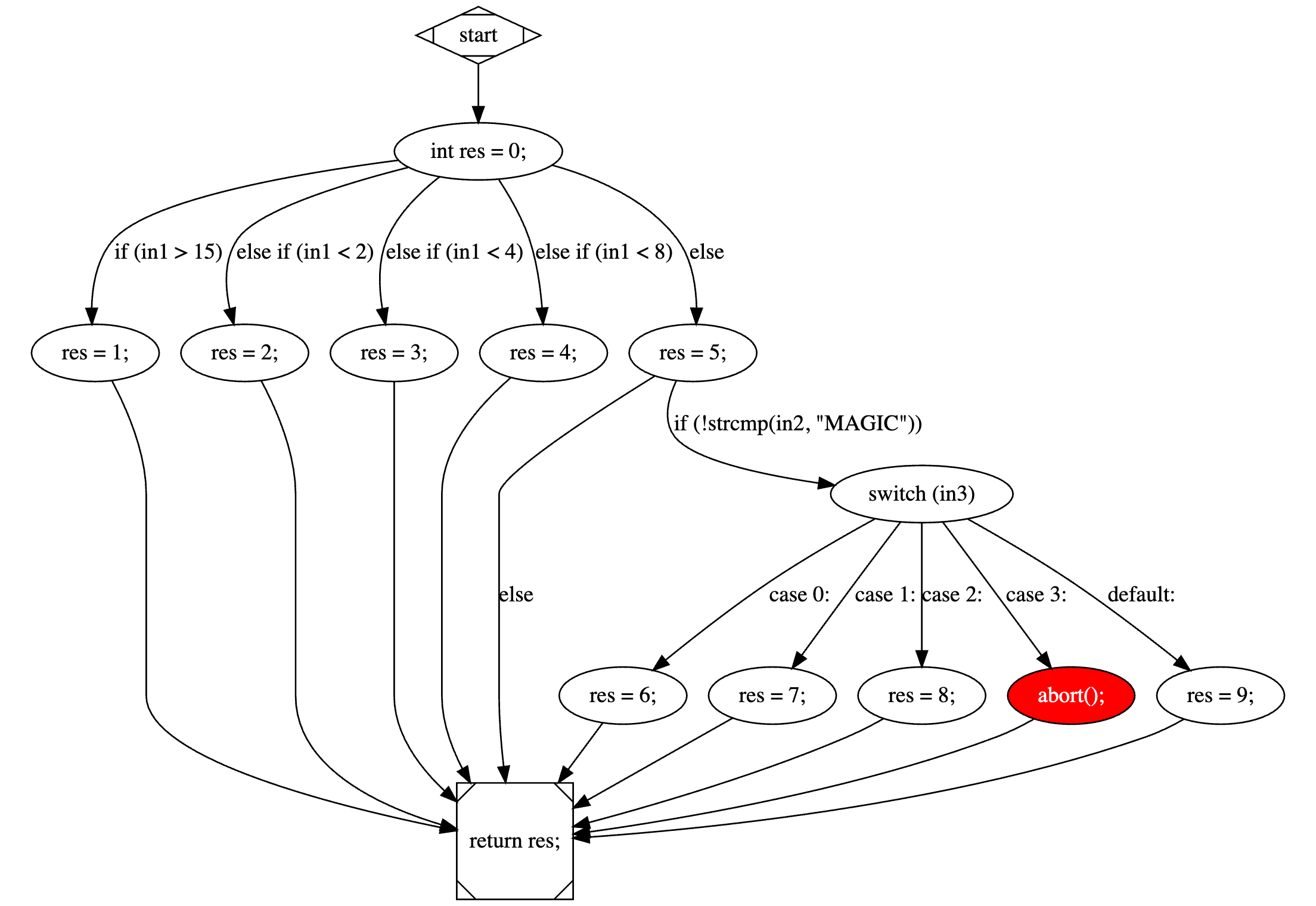}
    \caption{Control flow graph for figure \ref{fig:sourcecode}. Note the crash at abort, colored red.}
    \label{fig:cfg_empty}
\end{figure}

Assuming the use of a typical coverage-guided grey-box fuzzer, let's suppose we first discover the branch leading from \verb|int res = 0;| to \verb|res = 1;|, followed by discovery of the branch to \verb|res = 2;|, perhaps by using an input that inserts the commonly occurring constant value 0 into variable \verb|in1|. 
We then go on to discover the other 3 possible blocks reachable from \verb|int res = 0;|.
The covered blocks are marked green in the following:

\begin{figure}[H]
    \centering
    \includegraphics[width=\linewidth]{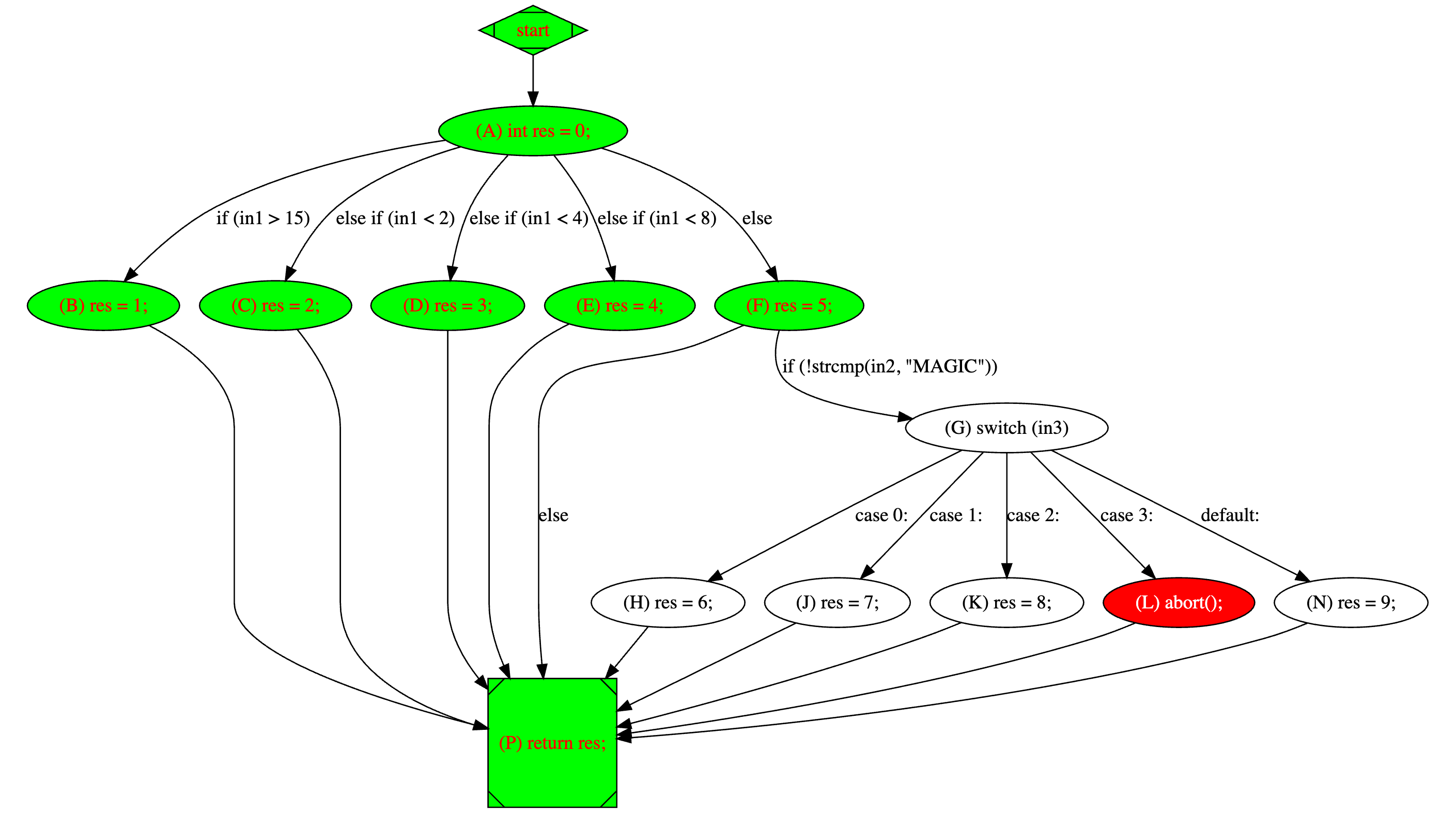}
    \caption{A control flow graph for the program in figure \ref{fig:sourcecode}, with covered blocks filled green.}
    \label{fig:cfg_covered}
\end{figure}

A common fuzzer strategy, employed in both LibAFL and AFL++ is to select a subset of the input corpus that achieves 100\% coverage of the currently discovered blocks; this subset is internally referred to as `favored'. 
It is selected by sorting the list of corpus entries in such a way that the fastest executing, shortest inputs (calculated by \texttt{execution\_time * input\_length}) are first; and then greedily selecting inputs from this sorted list until all current covered blocks are represented.
Favored inputs are more likely to be selected for mutation; in this case, all 5 inputs would be marked favored (as all 5 are required to cover all edges discovered), which is reasonable.
Additional gains can be achieved by weighting the selection probability for each input in the corpus.
This is what \ourFuzzer{} does, but using a novel weighting algorithm explained in the remainder of this section.

\subsection{Background Concepts}
\label{sec:formal}

We assume that at a given point in the fuzzer's campaign, the minimum (dynamic, updated) state of the fuzzer that we \emph{must} consider is the set of inputs in the corpus and the set of basic blocks that have been covered to date. 
There are further elements of the state that could be considered, for instance the exact priority assigned to elements of the input corpus by the scheduler, but these two are sufficient dynamic information to define the underlying idea of our approach. 

This relies in turn on two ideas: The set of uncovered basic blocks in the CFG that are reachable from a given covered basic block, and the set of minimal length, loop free paths where the first basic block in the path has already been covered and all subsequent blocks in the path are uncovered. 
These represent the available fresh territory that can be reached from that discovered already. 
The CFG is the map of the territory and the executions of the target SUT are the means by which we explore this map.

In the following we provide a formal definition of the concepts used by \ourFuzzer{}. 
Illustrative explanations of some of these concepts are provided in subsequent subsections.

Let \( \text{CFG} \) denote a control flow graph of a given program \( P \), defined as a tuple \( \text{CFG} = (BB, Edges) \), where:
\begin{itemize}
    \item \( BB \) is the set of basic blocks in the CFG. A basic block is defined as a straight line code section containing no branches (except for any incoming branches at the beginning and any outgoing branches at the end).
    \item \( Edges \subseteq BB \times BB \) represents the set of transitions between blocks, with each transition indicating possible control flow from one block to another. Note that the control flow graph is a directed graph.
\end{itemize}

Given a specific input, we assume that executing the program with this input will always result in the same set of blocks being covered. 
\begin{itemize}
    \item \( I \) is the complete input space
    \item \( i \in I \) is an input belonging to the input space
    \item \( P(i) \) is the path in the CFG obtained by executing the program \( P \) on input \( i \). If the input has been executed, the basic blocks in the path are termed \emph{covered}.
    \item \( Cov(i) \subseteq BB \) is the set of basic blocks in $P(i)$
\end{itemize}

For a fuzzing campaign input corpus \( C, C \subseteq I \), define:

\begin{itemize}
    \item \( AllCov(C) = \bigcup_{i \in C} Cov(i) \) \vspace{0.2cm}
    \item $DU\!N$ is short for Direct Uncovered Neighbour: 
    \begin{equation*}
        \begin{split}
            DU\!N(bb) = \{ x \in B\!B ~|~ \exists Edge(bb, x) ~\land~ bb \in B\!B ~\land~ \\ 
            Edge \in Edges(bb, x) ~\land~ x \notin AllCov(C) \} 
        \end{split} 
    \end{equation*}
    \item \( ADU\!N \) is short for All Direct Uncovered Neighbours.
    \begin{equation*}
        ADU\!N(Cov(i)) = \{ DU\!N(bb) ~|~ bb \in Cov(i) \}
    \end{equation*}
\end{itemize}

$ADU\!N(Cov(i))$ is then the set of uncovered basic blocks that are one-step reachable from the set of blocks in the CFG that correspond to the execution path for input $i$.
\vspace{3mm}

Let \( R(start, end) \subseteq Edges \) be a minimal length sequence of transitions (considered as a Route) in the control flow graph to get from \( start \) to \( end \). Overload $R(start, end)$ to mean also the set of basic blocks in the route.
Assume \( R(start, end) \) is a non-empty ordered pair with the left basic block the starting point and the right basic block the end point, with individual transitions between them represented as ordered pairs \( (bb_i, bb_{i+1}) \). Define:
\begin{itemize}
    \item \( start \) as a specific starting block \( \in B\!B \),
    \item \( end \) as a specific ending block \( \in B\!B \).
\end{itemize}

Then, \( R(start, end) \) must satisfy the following conditions:
\begin{enumerate}
    \item The first transition in \( R(start, end) \) must start from \( start \), i.e., if \( (bb_1, bb_2) \) is the first element of \( R(start, end) \), then \( bb_1 = start \).
    \item Each subsequent transition must start at the end of the previous transition, and there are no cycles.
    \item The last transition in \( R(start, end) \) must end at \( end \), i.e., if \( (bb_k, bb_{k+1}) \) is the last element of \( R(start, end) \), then \( bb_{k+1} = end \).
\end{enumerate}
As above, we overload $R(start, end)$ to stand for the set of blocks traversed on the route, as well as the set of edges on the path, the difference being clear from context.

Now define:
\begin{itemize}
    \item \( RU\!B \) is short for Reachable Uncovered Blocks. These are the uncovered basic blocks that can be reached from a given covered block, $bb$, in some transition sequence of uncovered blocks.
    \begin{equation*}
        \begin{split}
        RU\!B(bb) = \Bigl\{ x \in B\!B ~|~ \exists R(bb, x) \land \forall (s,e) \mbox{ in } R(bb,x), \\
        \{ s, e \} \cap (AllCov(C) \setminus \{bb\}) = \emptyset \Bigr\}
        \end{split}
    \end{equation*}
    \item \(ARU\!B \text{ is short for All Reachable Uncovered Blocks.} \\
    \mbox{Let } X \subseteq Cov(C), \mbox{ then }
    ARU\!B(X) = \bigcup_{bb \in X} RU\!B(bb) \)
\end{itemize}

\subsection{Direct Uncovered Neighbours}
\label{sec:direct_neighbours}

The intuition behind our approach is to use the knowledge of the control flow graph to select inputs for mutation, prioritising based on the number of uncovered blocks that can be reached in the CFG from the execution path for the input.
We define a \emph{direct uncovered neighbour} as being an uncovered  block that is reachable from the current path by following a single edge, but is not in the set of blocks covered by any input in the corpus.

Taking for example the control flow graph shown in figure \ref{fig:cfg_covered}.
If we have an input covering the edges $\{A, B, P\}$, we see that from A it is possible to reach $\{B, C, D, E, F\}$, and from $B$ it is possible to reach only $P$.
So we have a set of \emph{possible} \emph{direct uncovered neighbours} of $\{B, C, D, E, F, P\}$.
If we first eliminate blocks that have been covered by this input, we are left with $\{C, D, E, F\}$.
If we also exclude the other covered edges marked in green -- $\{C, D, E, F\}$ -- we are left with the empty set $\{\}$.

If instead we take an input covering $\{A, F, P\}$, we have a set of \emph{possible} \emph{direct uncovered neighbours} $\{B, C, D, E, F, G, P\}$.
Eliminating the blocks that have been covered leaves us with one \emph{direct uncovered neighbour}: ${G}$.

A naive approach would be to give a weighting equal to the number of \emph{direct uncovered neighbours}, which from the current state leaves all paths apart from $\{A, F, P\}$ with a weight of 0, and $\{A, F, P\}$ itself with a weight of 1.
This would mean that we would only ever select the latter input (covering $\{A, F, P\}$) for mutation.
Hopefully it is intuitive why this is a good approach for achieving more coverage; and a fuzzer cannot find bugs in code that it does not cover.

\subsection{Reachable Blocks}
\label{sec:reachable}

We extend the concept of \emph{direct uncovered neighbours}, for an input $i$, to be the set of any blocks that are reachable from the set of blocks covered when executing $i$, without needing to traverse any blocks that have already been covered by other inputs during the fuzzing process --- we call this set the \emph{reachable blocks}.

To find the set of \emph{reachable blocks} from a given set of blocks covered by executing an input $i$, we perform a breadth-first search starting with each of the blocks covered by $i$ added to the queue, and the complete set of blocks covered by \emph{all} inputs added to visited.
We keep track of the depth at which each \emph{reachable block} was visited.
Pseudocode describing this process is as follows:

\noindent
\begin{minipage}{\linewidth}
\begin{lstlisting}[morekeywords={func,for,in,not,while,return}]
func calc_reachable_blocks(fuzzer, input):
    queue = Queue()
    depth = 0
    covered = fuzzer.execute(input).covered_blocks
    for block in covered:
        queue.push_back( (block, depth) )
    
    visited = HashSet()
    for block in fuzzer.all_covered_blocks:
        visited.insert(block)
    
    reachable_blocks = []
    
    while not queue.is_empty():
        (block, depth) = queue.pop_front()
        for succ in block.successors:
            if not visited.contains(succ):
                visited.insert(succ)
                reachable_blocks.append((succ, depth + 1))
                queue.push_back( (succ, depth + 1) )
    
    return reachable_blocks
\end{lstlisting}
\end{minipage}

\noindent
For the example program in Figure \ref{fig:sourcecode}, with input covering $\{A, F, P\}$, starting with \\\texttt{fuzzer.all\_covered\_blocks} being \texttt{[A, B, C, D, E, F, P]} the resultant value of \texttt{reachable\_blocks} is \texttt{[(G, 1), (H, 2), (J, 2), (K, 2), (L, 2), (N, 2)]}.

\subsection{Rarity Weighting}
\label{sec:rarity}

\begin{figure}[t]
    \centering
    \includegraphics[width=0.8\linewidth]{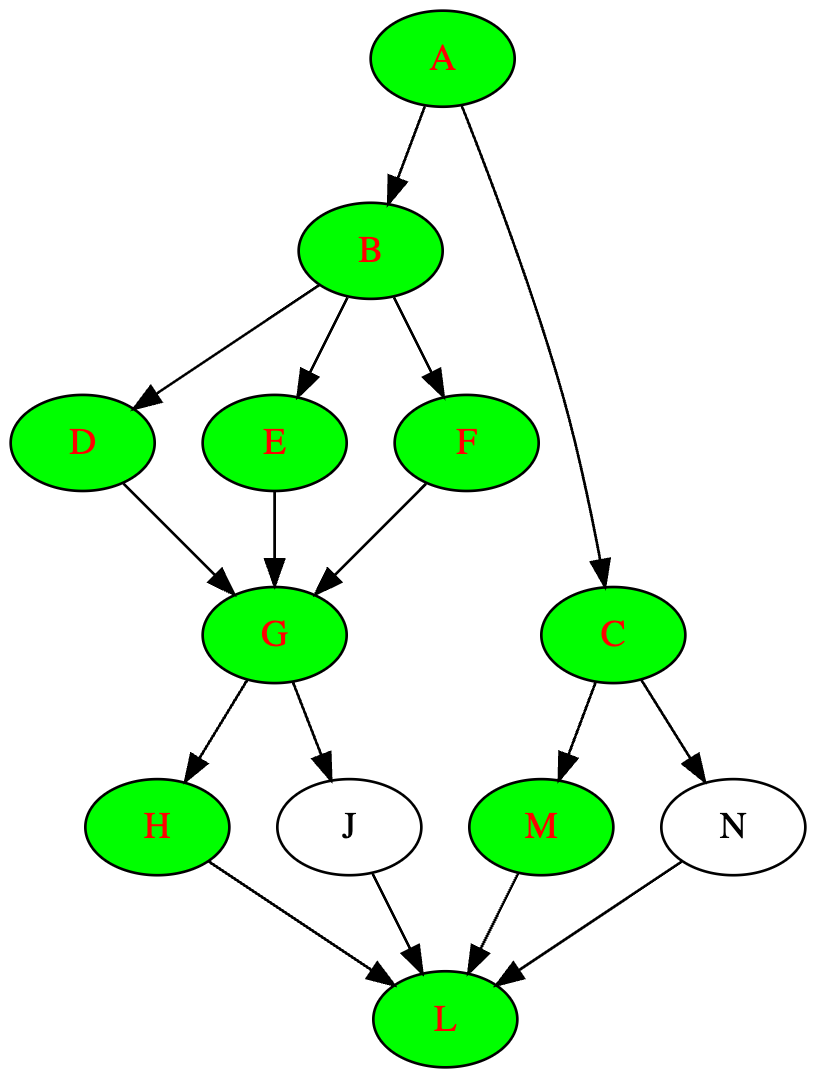}
    \caption{A control flow graph for a program, with covered blocks filled green.}
    \label{fig:graph_rarity}
\end{figure}

We use the rarity weighting to attempt to equalise the amount of fuzzing effort that is put towards reaching each \emph{reachable block}. 
Take for example the CFG shown in Figure \ref{fig:graph_rarity}.
Again, we use a green fill to indicate nodes that have been covered.
Let's say we have inputs in the fuzzing corpus that cover the following paths, and \emph{reachable blocks}:

\noindent
\begin{minipage}{\linewidth}
\begin{verbatim}
Path                  Reachable Blocks
[A, B, D, G, H, L]    {(J, 1)}
[A, B, E, G, H, L]    {(J, 1)}
[A, B, F, G, H, L]    {(J, 1)}
[A, C, M, L]          {(N, 1)}
\end{verbatim}
\end{minipage}
\vspace{2mm}

Directly taking the number of reachable blocks as the weighting, every one of these inputs would have a score of 1.
This means that 3 out of 4 times we would select an input for mutation that has J as reachable.
There is no particular reason that we should spend 3 times the amount of effort on attempting to reach that block as opposed to N.
Thus we propose to normalise the effort, by keeping track of the number of times each block appears in the sets of \emph{reachable blocks} and giving a score proportional to the inverse of this as follows:

\noindent
\begin{minipage}{\linewidth}
\begin{lstlisting}[language=pseudo]
reachability_freq = {} # empty dictionary
# Calc the number of times we see each reachability
for input in fuzzer.corpus:
    reachable = calc_reachable_blocks(fuzzer, input)
    for (block, depth) in reachable:
        freq = reachability_freq[(block, depth)]
        if freq is None:
            freq = 0
        reachability_freq[(block, depth)] = freq + 1

for input in fuzzer.corpus:
    # Here, weighting is the probability of an input 
    # being selected for mutation
    weighting = 0
    reachable = calc_reachable_blocks(fuzzer, input)
    for (block, depth) in reachable:
        freq = reachability_freq[(block, depth)]
        weighting += 1 / freq
\end{lstlisting}
\end{minipage}

\noindent
\begin{minipage}{\linewidth}    
\noindent Using this new calculation, we get the following:

\begin{small}
\begin{verbatim}
reachability_freq = {
    (J, 1): 3,
    (N, 1): 1
}

Path                Reachable Blocks  Weighting
[A, B, D, G, H, L]  {(J, 1)}          1 / 3
[A, B, E, G, H, L]  {(J, 1)}          1 / 3
[A, B, F, G, H, L]  {(J, 1)}          1 / 3
[A, C, M, L]        {(N, 1)}          1 / 1
\end{verbatim}
\end{small}

\noindent Using these weightings, we now have an equal probability of selecting an input for mutation that borders either J or N.
\end{minipage}

\subsection{Depth Weighting}
\label{sec:depth}

Additionally, we weight the \emph{reachable blocks} based on their \emph{depth}, which is the minimum number of edges that need traversing to reach the given block.
Again, we use an inverse for the weighting: 1 $\div$ depth.
Note that the instrumentation we use does not instrument dominated or post-dominated blocks, thus in practice the depth is actually the number of conditionals that need to be passed to reach a given block.
Figure \ref{fig:graph_minimised} illustrates the blocks that are actually instrumented after this minimisation step; these nodes are colored blue:

\begin{figure}[H]
    \centering
    \includegraphics[width=0.8\linewidth]{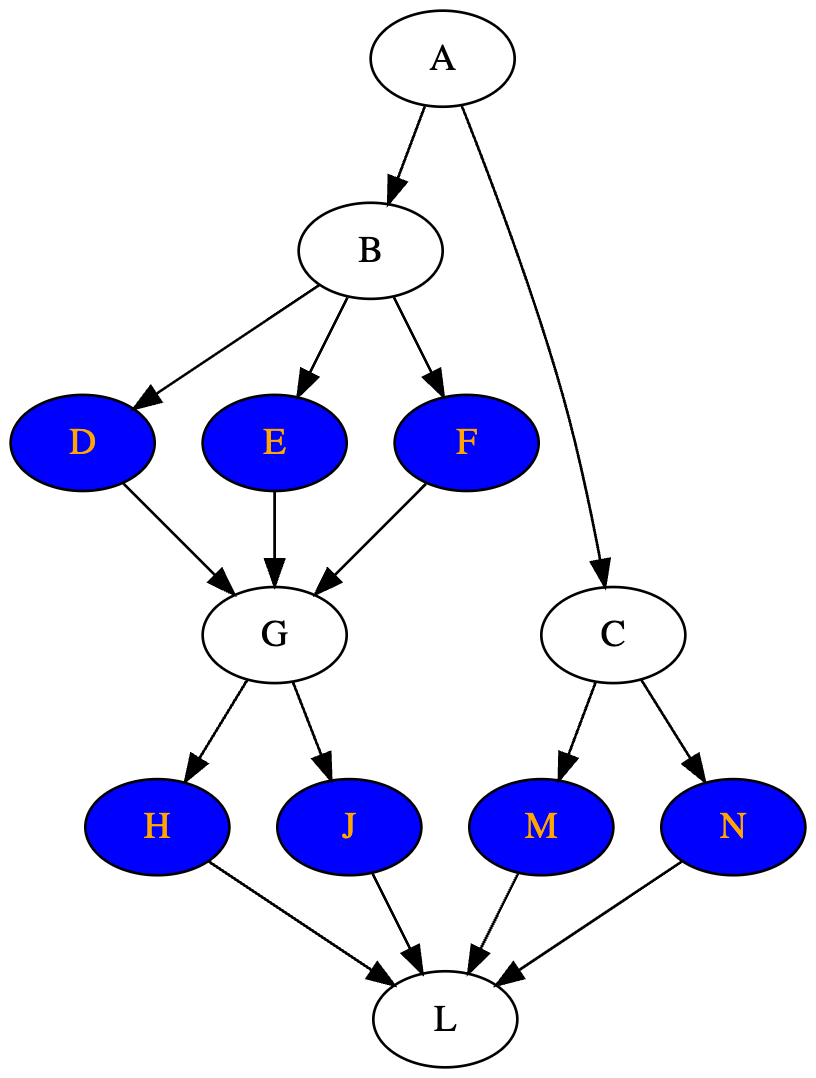}
    \caption{A control flow graph showing the minimal set of nodes (filled blue) that need instrumenting in order to infer the complete set of nodes covered by a program execution}
    \label{fig:graph_minimised}
\end{figure}

\subsection{Weighted Fuzzer Corpus Scheduling}
\label{sec:prescientSchduler}

Combining all of the elements described previously, we get the following score calculation:

\noindent
\begin{minipage}{\linewidth}
\begin{lstlisting}[language=pseudo]
func compute_score(reachability_freq, input):
   reachability_score = 0
   reachable = calc_reachable_blocks(fuzzer, input)
   for (block, depth) in reachable:
      # Using inverse depth prioritises more 
      # immediately reachable blocks
      score = 1 / depth
      # if many inputs have the same reachability, 
      # weight it less; thus inputs with rarely seen 
      # reachable_blocks are more likely to be chosen
      score *= 1 / reachability_freq[(block, depth)]
      reachability_score += score
   
   # input.exec_time() is the amount of time taken to 
   # execute this input, slower inputs get a lower 
   # score
   time_score = 1 / input.exec_time()
   
   return reachability_score * time_score
\end{lstlisting}
\end{minipage}

\noindent
\begin{minipage}{\linewidth}
\begin{lstlisting}[language=pseudo]
func compute_all_scores(fuzzer):
    reachability_freq = {}
    # Compute how many times we see each reachability
    for input in fuzzer.corpus:
        reachable = calc_reachable_blocks(fuzzer, input)
        for (block, depth) in reachable:
            freq = reachability_freq[(block, depth)]
            if freq is None:
                freq = 0
            reachability_freq[(block, depth)] = freq + 1

    input_weightings = {}
    for input in fuzzer.corpus:
        score = compute_score(reachability_freq, input)
        input_weightings.insert(input, score)

    return input_weightings
\end{lstlisting}
\end{minipage}

Note that because the initial set of covered blocks (\verb|fuzzer.all_covered_blocks| in \verb|calc_reachable_blocks|) changes every time that new coverage is found (i.e. any new coverage is \emph{added} to it), the complete set of scores must be recalculated whenever a new entry discovering coverage is added to the corpus.
For some of the benchmarks that we evaluated on, there were over 20,000 basic blocks and over 1,000 inputs in the corpus; in these cases \\\verb|compute_all_scores| has excessive overhead.

In order to mitigate this, we store the amount of time that \verb|compute_all_scores| takes to run, and only allow it to be run again once 10$\times$ that amount of time has elapsed (we refer to this as a \emph{cooldown period}).
If new coverage is discovered in the meantime, a flag is set to indicate that the scores need recomputing; this flag is checked whenever a new input is to be selected for mutation.
At the beginning of the fuzzing campaign, it is possible that many inputs achieving new coverage are discovered during the mutation of a single seed -- the scores are only used for selecting the next input for fuzzing, hence do not need recalculating until we actually go to select a new input.

During the \emph{cooldown period}, any new inputs that are added to the corpus are assigned a score equal to the average of all other inputs in the corpus.
We see that at the beginning of a fuzzing campaign there are typically many inputs discovered during the cooldown period, but afterwards the discovery rate slows enough that each new discovery results in an immediate recalculation when the next corpus entry is selected.

The most similar approach, FishFuzz, pre-computes a static distance between functions and they give the argument that fully computing basic block distances would be too costly; we have not found this to be the case with the implementation of the \emph{cooldown period}.
Our dynamic approach has a key advantage in that indirect branches can be resolved at runtime by looking at the coverage feedback for each input; indirect branching is common in object-oriented languages, where dynamic dispatch is used.
A further result of using dynamic computations, is that we could alternate between directed grey-box fuzzing targets at runtime if so desired; whereas this would typically require costly re-compilation of the SUT.
In some ways, we have unintentionally built a \emph{universal} directed grey-box fuzzer.

\section{Implementation}
\label{sec:impl}

In this section, we describe the implementation specific details of our approach as it is used in \ourFuzzer.
We found that storing the complete control flow graph (CFG), in such a way as to be able to align it with the coverage feedback received from individual executions, to be a complex endeavour; we believe that this is the reason that our relatively simple idea has not been done already.
We implemented a version of the technique using LibAFL \cite{fioraldi2022libafl}, and use a custom LLVM compiler pass in order to store the CFG details to a file as part of the SUT's compilation process.
This file is read in at runtime by the fuzzer.
Note that as this is a more general technique, it is not built on any of the work of the previous chapters, which were specifically aimed at detection information leaks.
That said, this technique could be adopted into NIFuzz and would likely improve its results.

In particular, this is done using a modified version of the \texttt{SanitizerCoverage} pass that is built in to LLVM; specifically we override the \texttt{trace-pc-guard} coverage instrumentation.
\texttt{trace-pc-guard} provides user-defined callback when the SUT is initialised, that gives pointers to the beginning and end of an array where each element maps to a particular basic block in the CFG.
There is an additional separate callback to a user-defined function when entering each block in the SUT's CFG; this callback provides (as an argument) a pointer to one of the elements of the aforementioned array.
It is expected that during the initialisation callback, the user will assign a unique value to each array element, so that during the latter callback they can establish which block has been entered.

Instead of leaving it to the user-defined initialisation callback to assign values to each edge, we preset these values during the compiler pass.
As we discovered, it is likely that the reason for requiring an initialisation callback to allow the user to assign unique values to each block at runtime, is because of the difficulty in keeping track of which values have already been assigned during the compilation process.
As it is common in a C/C++ build to compile each file to a \verb|.o| object file separately, and then link these later, some state must be retained by the compiler between invocations.
We overcame this by storing a counter in the CFG output file that we were producing to pass in to the fuzzer, and beginning the block enumeration at the value following on from that last assigned.
Note that it is possible to register a link-time optimisation (LTO) pass with LLVM, at which point all edges can be labelled in a single pass; unfortunately this severely limits the LLVM versions that can be simultaneously supported, so we chose not to do this.

Now we come to the file format for the CFG file itself.
Firstly, we discuss the required information for each edge in the CFG.
During the compiler pass, we iterate through all functions, and within these each basic block, and finally each instruction within the basic block.
As shown in Figure \ref{fig:graph_minimised}, \texttt{trace-pc-guard} does not instrument every block in the CFG; only the minimal subset required in order to determine the exact set of covered blocks.
We chose to include all blocks in our CFG file representation, whether assigned an identifier or not; for this reason, we assign each block a unique identifier that is separate to it's index in the coverage map.

For each basic block, we therefore store: a unique block identifier, the coverage map index (if one has been assigned), a list of functions (names) called within the block, a list of the unique identifiers for the \emph{successor} blocks, and the number of indirect function calls.
The list of function calls is populated whilst iterating through the instructions that make up the block; whilst we could attempt to directly list the entry block of the function as a \emph{successor}, definitions that are in other files cannot be resolved yet.
Additionally, we found that being able to determine when the fuzzer has entered a new function by simply looking at the list of called functions to be a useful debugging aid.
We do not know the target of indirect function calls at compilation time, however we track these and in our reachability calculations assume that an indirect function call discovers one new basic block.
This may not be the case in practice, as the indirect call may target a function that has already been explored, but in SUTs that make use of indirection this was a better approximation than ignoring them.
During the LLVM pass, one can determine whether a \texttt{CallInst} is indirect using the \texttt{IsIndirectCall()} member function.

For each function, we store: the function name and a list of the unique identifiers for each basic block within the function.
A pseudocode description of the pass is shown in Figure \ref{fig:instr_pass}.


\begin{figure}[t]
    \begin{lstlisting}[language=pseudo]
struct BBInfo {
    ptr: BasicBlock *,
    uid: int,
    coverage_map_index: int,
    called_funcs: String[],
    successor_bbs: BasicBlock *[],
    num_indirect_calls: int
}

latest_coverage_map_index, lastest_block_uid = \
    fetch_from_CFG_file()
bbs_in_function_named = {}
bb_infos = []

for function in module:
    bbs_in_func: BBInfo[] = []
    for bb in function:
        bb_info = BBInfo()

        for succ in bb.successors:
            bb_info.successor_bbs.append(succ)

        bb_info.uid = latest_block_uid
        latest_block_uid += 1
        
        if bb.needs_instrumenting:
            bb.insert_sancov_callbacks()
            bb_info.coverage_map_index = \
                latest_coverage_map_index
            latest_coverage_map_index += 1
        else:
            bb_info.coverage_map_index = -1
        
        for inst in bb:
            if inst.isFuncCall():
                if inst.isIndirectCall():
                    bb_info.num_indirect_calls += 1
                else:
                    func_name = inst \
                        .get_called_function().name()
                    bb_info.called_funcs.append(func_name)
        bbs_in_func.append(bb_info)

    func_name = function.name()
    bbs_in_function_named[func_name] = bbs_in_func

update_CFG_file(
    latest_coverage_map_index, latest_block_uid, 
    bb_infos, bbs_in_function_named
)
    \end{lstlisting}
    \caption{A pseudocode description of our compiler pass}
    \label{fig:instr_pass}
\end{figure}

The CFG file is updated rather than overwritten at each stage; thus before starting the build process, the file should be cleared by the user if it has unwanted contents.
Additionally, as build systems such as \texttt{make} typically compile object files in parallel, we use a lock-file (the file system equivalent to a mutex) to ensure that there is no contention over \texttt{latest\_coverage\_map\_index}; that is, the lock is acquired in \texttt{fetch\_from\_CFG\_file} and released at the end of \texttt{update\_CFG\_file}.
The file contents themselves are essentially a binary serialisation of the arguments provided to the \texttt{update\_CFG\_file} function call.

\section{CFG File Parsing and CFG Reconstruction}

Our fuzzer takes, as an argument, a path to the CFG file.
Firstly, the file is deserialised in order to reconstruct the original data structures.
From here, we simplify the CFG, and cache as much information as possible to aid our reachability algorithm during the fuzzing campaign -- as this needs to be ran potentially millions of times, and is computationally expensive.

In a first pass, we compute the set of directly reachable blocks from the entry of each function; we cache these so that they do not need to be recomputed every time a function call is encountered.
Next, for each basic block that has been instrumented with a callback, we compute and cache the set of directly reachable blocks that have also been instrumented, this includes any blocks within function calls made on the way.
This saves us from needing to traverse the uninstrumented blocks later on during the reachability search.
Finally, as we assign \texttt{uid}s consecutively, we store the basic block information into an array, where the index at which it is stored corresponds to the \texttt{uid}; which saves us a level of indirection when traversing the graph.

One additional issue to overcome, is the multiple definition of functions.
As our compiler pass does not run at link-time -- when all definitions would have been resolved -- we find that for some SUTs there are multiple definitions of some functions.
We could attempt to statically walk the CFG and resolve these, however any functions reachable only by indirect calls would be tricky (if not impossible) to resolve.
Instead we store lists of all definitions each function, then assume that the first in the list is the one used; if we later receive coverage feedback that indicates this was incorrect, we swap the actual encountered definition to be first in the list. 

One unplanned benefit of not using the link-time pass to create the CFG is that we also receive CFG information for shared libraries, which would otherwise be excluded as they are compiled as separate modules.

\section{General Fuzzer Implementation}

The fuzzer itself is modified from LibAFL's \emph{fuzzbench} fuzzer (note that while LibAFL is a library of fuzzing components, it also includes some example fuzzers), and as such inherits the included state-of-the-art functionality based on RedQueen \cite{aschermann2019redqueen} and MOpt \cite{lyu2019mopt}.
Unlike LibAFL's original \emph{fuzzbench} fuzzer, we do not use the corpus scheduler based on AFLFast \cite{bohme2016coverage}, but our own as described in section \ref{sec:prescientSchduler}.

\section{Evaluation}
\label{prescient:sec:eval}

In order to evaluate \ourFuzzer{}, we aimed to answer the following research questions:

\noindent \textbf{RQ1:} How does the coverage achieved by \ourFuzzer{} compare to other state-of-the-art fuzzers?

\noindent \textbf{RQ2:} How does the proposed scheduler compare to random scheduling and state-of-the-art weighted scheduling?

\noindent \textbf{RQ3:} How do the different weighting mechanisms described in Sections \ref{sec:direct_neighbours}, \ref{sec:reachable}, \ref{sec:rarity} and \ref{sec:depth} affect the rate of coverage discovery?

To answer these questions, we evaluated \ourFuzzer{} on the standard FuzzBench coverage benchmark suite \cite{metzman2021fuzzbench}. 
It is a collection of 22 real-world programs (and 1 artificial program -- \texttt{bloaty\_fuzz\_target}), and the benchmarks offer significant enough diversity that results generalise well.
More in depth details can be found in Table \ref{table:fuzzbench_programs}.
Under standard setup parameters, each program is run 20 times for 23 hours by each fuzzer being evaluated; to save on running costs, instances can be pre-empted, so in many cases less than 20 runs will complete the full 23 hours.
The service is generously maintained and operated free-of-charge by Google, and all fuzzers are subject to the same standards.

\begin{table}[]
\begin{footnotesize}
\begin{tabular}{llll}
\textbf{Benchmark}             & \textbf{dict} & \textbf{\# seeds} & \textbf{\# edges} \\
\hline \rowcolor[HTML]{EFEFEF} \texttt{bloaty\_fuzz\_target}           & false                    & 94                             & 89,530                           \\
\texttt{curl\_curl\_fuzzer\_http}       & false                    & 31                             & 62,523                           \\
\rowcolor[HTML]{EFEFEF} \texttt{freetype2\_ftfuzzer}                 & false                    & 2                              & 19,056                           \\
\texttt{harfbuzz-hb-shape-fuzzer}                 & false                    & 58                             & 10,021                           \\
\rowcolor[HTML]{EFEFEF} \texttt{jsoncpp\_jsoncpp\_fuzzer}       & true                     & 0                              & 5,536                            \\
\texttt{lcms\_cms\_transform\_fuzzer}                & true                     & 1                              & 6,959                            \\
\rowcolor[HTML]{EFEFEF} \texttt{libjpeg-turbo\_libjpeg\_turbo...}          & false                    & 1                              & 9,586                            \\
\texttt{libpcap\_fuzz\_both}            & false                    & 0                              & 8,149                            \\
\rowcolor[HTML]{EFEFEF} \texttt{libpng\_libpng\_read\_fuzzer}                  & true                     & 1                              & 2,991                            \\
\texttt{libxml2\_xml}                 & true                     & 0                              & 50,461                           \\
\rowcolor[HTML]{EFEFEF} \texttt{libxslt\_xpath}                 & true                     & 112                              & 34,860                           \\
\texttt{mbedtls\_fuzz\_dtlsclient}      & false                    & 1                              & 10,942                           \\
\rowcolor[HTML]{EFEFEF} \texttt{openssl\_x509}                  & true                     & 2,241                           & 45,989                           \\
\texttt{openh264\_decoder\_fuzzer}          & false                    & 1                              & 17,443                           \\
\rowcolor[HTML]{EFEFEF} \texttt{openthread\_ot-ip6-send-fuzzer}          & false                    & 0                              & 17,932                           \\
\texttt{proj4\_proj\_crs\_to\_crs\_fuzzer}               & true                     & 44                             & 6,156                            \\
\rowcolor[HTML]{EFEFEF} \texttt{re2\_fuzzer}                 & true                     & 0                              & 6,547                            \\
\texttt{sqlite3\_ossfuzz}               & true                     & 1,258                           & 45,136                           \\
\rowcolor[HTML]{EFEFEF} \texttt{stb\_stbi\_read\_fuzzer}        & true                     & 166                           & 5,026                           \\
\texttt{systemd\_fuzz-link-parser}      & false                    & 6                              & 53,453                           \\
\rowcolor[HTML]{EFEFEF} \texttt{vorbis\_decode\_fuzzer}              & false                    & 1                              & 5,022                            \\
\texttt{woff2\_convert\_woff2ttf\_fuzzer}               & false                    & 62                             & 10,923                           \\
\rowcolor[HTML]{EFEFEF} \texttt{zlib\_zlib\_uncompress\_fuzzer} & false                    & 0                              & 875                            
\end{tabular}
\end{footnotesize}

\caption{Statistics for the 23 programs that make up the default FuzzBench coverage benchmark suite. Note that the `dict' column indicates whether the program comes with a dictionary which helps with input generation.}
\label{table:fuzzbench_programs}
\end{table}

\subsection*{RQ1: How does the coverage achieved by \ourFuzzer{} compare to other state-of-the-art fuzzers?}

As the unique methods implemented by \ourFuzzer{} aim to improve input scheduling, and not the mutation engine itself, we would expect any gains in program coverage to come early in the fuzzing campaign.
Table \ref{tab:results} shows the relative performance for a set of fuzzers; the numerical value is the median code coverage percentage of the fuzzer across all benchmarks (relative to the best performing fuzzer on each -- thus a fuzzer that performed best on all benchmarks would score 100).
Note that the displayed results are made up the top-5 set of `default' fuzzers; these are fuzzers that are actively maintained and represent the state-of-the-art.
One of the big advantages of using FuzzBench is that each fuzzer has been setup by their maintainers, hence are more likely to be optimal setups.
To provide some context, aflplusplus (AFL++) can build with many clang or GCC versions, but a very specific setup including the clang linker and archiver are required to get maximum performance from it -- in an independent evaluation it is likely that someone would miss this and produce non-representative results.
Notably, FishFuzz is missing from this evaluation as there is no provided setup for FuzzBench; though given that it targets bugs that are detectable by sanitizers above all else, it is not clear whether it would compare favourably on coverage metrics.

\newcommand{\greenup}{\textcolor{green}{$\uparrow$}}
\newcommand{\reddown}{\textcolor{red}{$\downarrow$}}

\begin{table}[H]
    \centering
    \begin{large}
    \begin{tabular}{l@{\hskip 5mm}l@{\hskip 5mm}l}
        \textbf{Fuzzer} & \textbf{2 hours} & \textbf{23 hours} \\ \hline
        PrescientFuzz & 98.49 & 99.05 \\
        libafl & 96.93 & 97.79 \\
        aflplusplus & 95.69 & 95.45 \\
        honggfuzz & 92.26 & 93.53 \\
        libfuzzer & 88.77 & 91.81 \\
        afl & 82.49 & 84.07
    \end{tabular}
    \end{large}
    \caption{Table showing the relative median coverage for all benchmarks (mean percentage) at 2 and 23 hours. Note that `libafl' is the default libafl-based fuzzer submitted to FuzzBench; it uses an AFLFast-based corpus scheduler.}
    \label{tab:results}
\end{table}

We have chosen to list results at the 2 hour mark, as this is where \ourFuzzer{} has the greatest advantage over `libafl' (AFLFast-based scheduling); and 23 hours as this was the end time of the campaigns.
As can be seen, PrescientFuzz compares favourably with the other fuzzers, and notably outperforms its parent libafl.
As expected, it does better early on, with a 1.6\% advantage over libafl at 2 hours, dropping to 1.3\% advantage by 23 hours.
As can be seen by the gap between libafl and aflplusplus, the gap between fuzzers can be relatively small.

\subsection*{RQ2: How does the proposed scheduler compare to random scheduling and state-of-the-art weighted scheduling?}

\begin{figure}[H]
    \centering
    \includegraphics[width=1\linewidth]{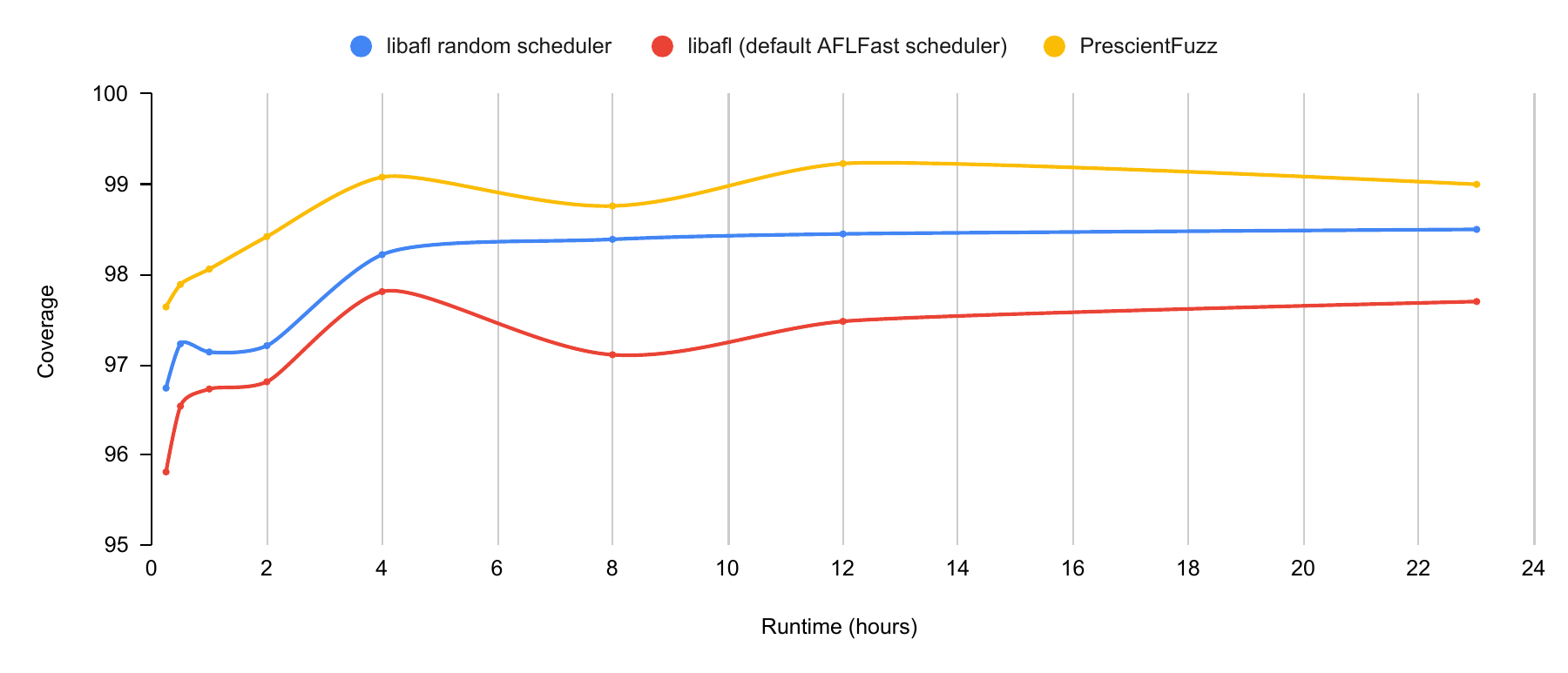}
    \caption{Chart showing the relative coverage for the libafl-based fuzzer setups.}
    \label{fig:libafl_comparison}
\end{figure}

Figure \ref{fig:libafl_comparison} shows the relative coverage scores over time for the various libafl-based fuzzers with more time granularity -- note that these aggregated statistics take a long time to compute, hence the sparsity of data points.

As can be seen, the random scheduler outperforms libafl's default AFLFast-based scheduling approach in the long run.
This was not something that we anticipated, but given that AFLFast is considered state-of-the-art for scheduling, it may mean that we have built the first scheduler to outperform random.
More detailed results broken down by benchmark can be found in Table \ref{table:schedulers_SUTs}.

\begin{table*}[t]
\begin{center}
\begin{tabular}{l|lll|ll}
\multicolumn{1}{c|}{\multirow{2}{*}{\textbf{Benchmark}}} & \multicolumn{3}{c|}{\textbf{Scheduler}}                            & \multicolumn{2}{c}{\textbf{Vargha Delaney}}                              \\
\multicolumn{1}{c|}{}                                    & \textbf{PrescientFuzz} & \textbf{AFLFast-based} & \textbf{Random} & \textbf{PF--Fast} & \textbf{PF--Rand} \\ \hline
\rowcolor[HTML]{EFEFEF} \texttt{bloaty\_fuzz\_target}                  & 6,284                          & 6,305.5                        & 6,346.5                 & 0.5                                                   & 0.31                                           \\
\texttt{curl\_curl\_fuzzer\_http}              & 10,748.5                       & 10,874                        & 10,854                  & 0.04                                                  & 0.12                                           \\
\rowcolor[HTML]{EFEFEF} \texttt{freetype2\_ftfuzzer}                   & 11,621.5                       & 11,225                         & 11,621.5                & 0.7                                                   & 0.5                                            \\
\texttt{harfbuzz\_hb-shape-fuzzer}             & 11,081                         & 11,074                         & 11,113                  & 0.55                                                  & 0.36                                           \\
\rowcolor[HTML]{EFEFEF} \texttt{jsoncpp\_jsoncpp\_fuzzer}              & 517                           & 517                           & 517                    & 0.62                                                  & 0.53                                           \\
\texttt{lcms\_cms\_transform\_fuzzer}          & 2,139.5                        & 2,076.5                        & 2,104                   & 0.77                                                  & 0.62                                           \\
\rowcolor[HTML]{EFEFEF} \texttt{libjpeg-turbo\_libjpeg\_t...} & 3,079                          & 3,078                          & 3,078                   & 0.77                                                  & 0.74                                           \\
\texttt{libpcap\_fuzz\_both}                   & 2,840                          & 2,770                          & 2,675                   & 0.75                                                  & 0.8                                            \\
\rowcolor[HTML]{EFEFEF} \texttt{libpng\_libpng\_read\_fuzzer}          & 1,999.5                        & 1,999                         & 1,998                   & 0.71                                                  & 0.61                                           \\
\texttt{libxml2\_xml}                          & 15,688                         & 15,625                         & 15,620                  & 0.85                                                  & 0.77                                           \\
\rowcolor[HTML]{EFEFEF} \texttt{libxslt\_xpath}                        & 11,125                         & 10,974                         & 11,018                  & 0.97                                                  & 0.91                                           \\
\texttt{mbedtls\_fuzz\_dtlsclient}             & 3,552                          & 3,105                          & 3,416.5                 & 0.82                                                  & 0.62                                           \\
\rowcolor[HTML]{EFEFEF} \texttt{openh264\_decoder\_fuzzer}             & 9,440                          & 9,446                          & 9,475                   & 0.47                                                  & 0.33                                           \\
\texttt{openssl\_x509}                         & 5,830                          & 5,823                          & 5,822                   & 0.78                                                  & 0.74                                           \\
\rowcolor[HTML]{EFEFEF} \texttt{openthread\_ot-ip6-send-fuzzer}        & 3,559                          & 3,548                          & 3,573                   & 0.64                                                  & 0.53                                           \\
\texttt{proj4\_proj\_crs\_to\_crs\_fuzzer}     & 7,427                          & 7,202.5                        & 7,399.5                 & 0.91                                                  & 0.63                                           \\
\rowcolor[HTML]{EFEFEF} \texttt{re2\_fuzzer}                           & 2,856                          & 2,856.5                        & 2,858                   & 0.51                                                  & 0.44                                           \\
\texttt{sqlite3\_ossfuzz}                      & 20,798                         & 20,714                         & 20,883.5                & 0.62                                                  & 0.32                                           \\
\rowcolor[HTML]{EFEFEF} \texttt{stb\_stbi\_read\_fuzzer}               & 2,197                          & 2,140.5                        & 2,187                   & 0.81                                                  & 0.76                                           \\
\texttt{systemd\_fuzz-link-parser}             & 239                           & 237                           & 237                    & 0.56                                                  & 0.58                                           \\
\rowcolor[HTML]{EFEFEF} \texttt{vorbis\_decode\_fuzzer}                & 1,252.5                        & 1,250                          & 1,253                   & 0.70                                                  & 0.49                                           \\
\texttt{woff2\_convert\_woff2tt\_fuzzer}       & 1,186.5                        & 1,184.5                        & 1,178.5                 & 0.64                                                  & 0.76                                           \\
\rowcolor[HTML]{EFEFEF} \texttt{zlib\_zlib\_uncompress\_fuzzer}        & 449.5                         & 450.5                         & 450                    & 0.42                                                  & 0.45                                          
\end{tabular}
\end{center}

\caption{Edge coverage broken down by program. Note that `median' here refers to the number of edges found in the median run (when sorted by coverage), and `AFLFast-based' refers to the default \texttt{libafl} scheduler. The `PF--Fast' column gives the Vargha-Delaney A12 measure between PrescientFuzz and AFLFast-based schedulers, and the `PF--Rand' column gives the same measure between PrescientFuzz and the Random scheduler.}
\label{table:schedulers_SUTs}
\end{table*}

\subsection*{RQ3: How do the different weighting mechanisms described in Sections \ref{sec:direct_neighbours}, \ref{sec:reachable}, \ref{sec:rarity} and \ref{sec:depth} affect the rate of coverage discovery?}

\begin{figure}[H]
    \centering
    \includegraphics[width=1\linewidth]{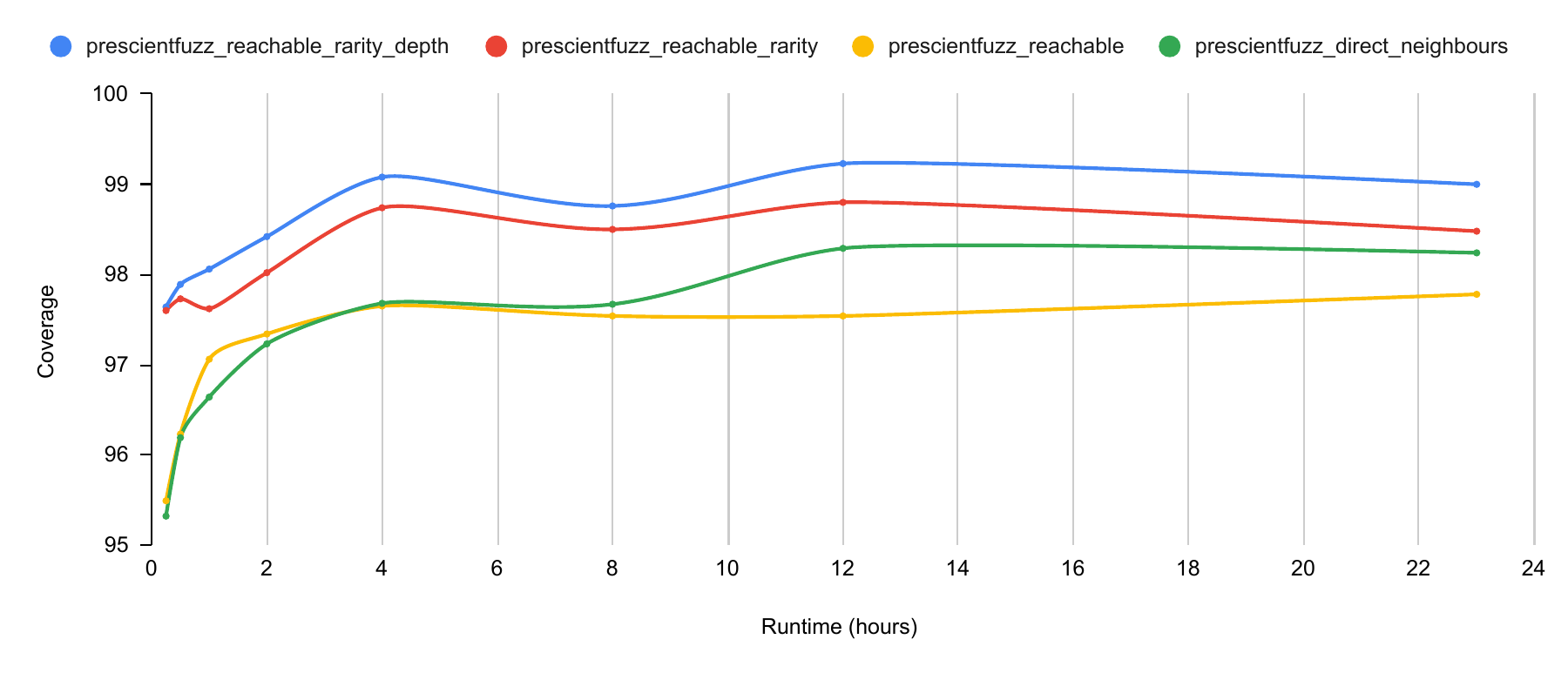}
    \caption{Chart showing the relative coverage achieved with different scheduling calculation steps included.}
    \label{fig:subschedulers_comparison}
\end{figure}

To answer this question, we compare the aggregated coverage for all of the different scheduling setups over time; this is shown in Figure \ref{fig:subschedulers_comparison}.
Here the fuzzer name for the various \texttt{prescientfuzz} setups indicates which features were enabled.
The \texttt{prescientfuzz\_direct\_neighbours} setup has scheduling based only on the number of direct uncovered neighbours, as described in section \ref{sec:direct_neighbours}.
The \texttt{prescientfuzz\_reachable*} setups have scheduling based on the total number of reachable uncovered blocks as described in section \ref{sec:reachable}.
The \texttt{prescientfuzz\_reachable\_rarity*} setups additionally weight the schedule based on rarity as described in section \ref{sec:rarity}, and finally \texttt{prescientfuzz\_reachable\_rarity\_depth} also uses the depth metric from section \ref{sec:depth} (this is referred to previously as just \texttt{PrescientFuzz} as it contains all of the features).

\section{Discussion}

\begin{figure}[t]
    \includegraphics[width=\linewidth]{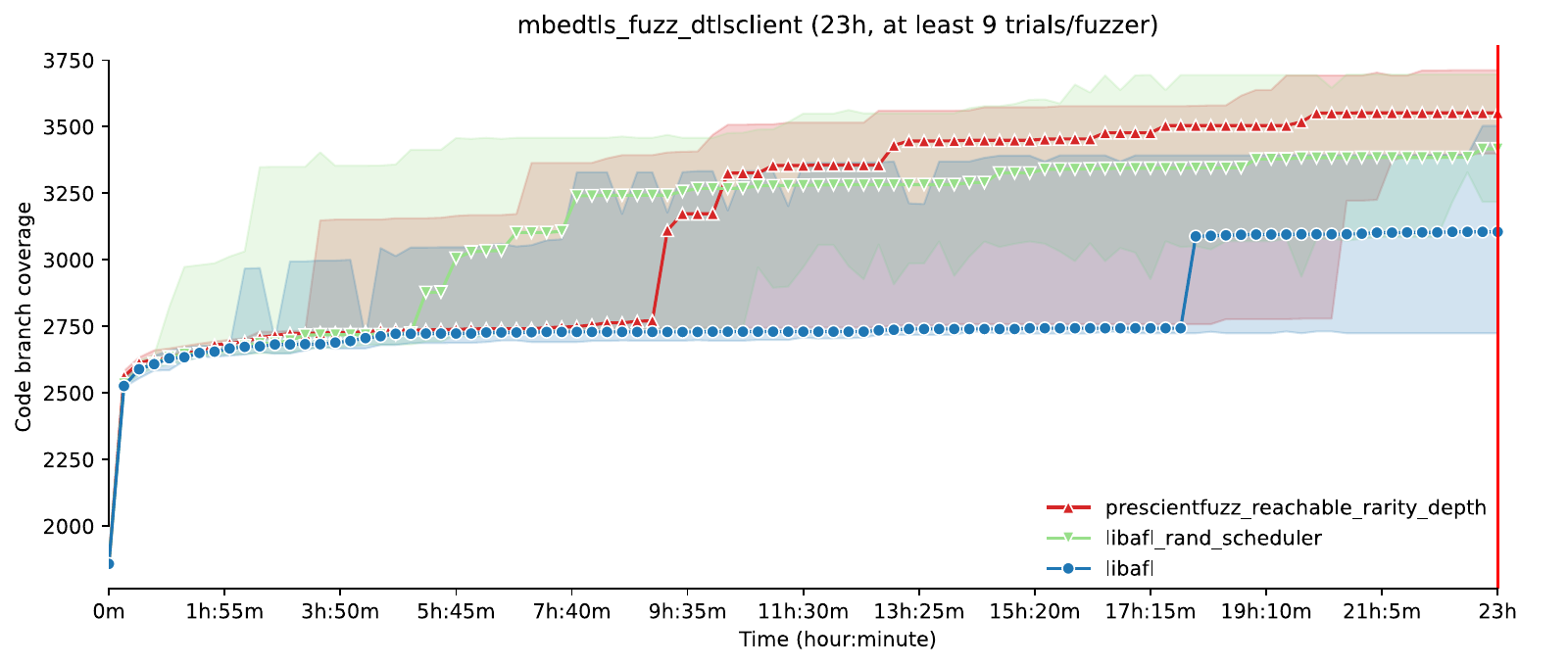}
    \includegraphics[width=\linewidth]{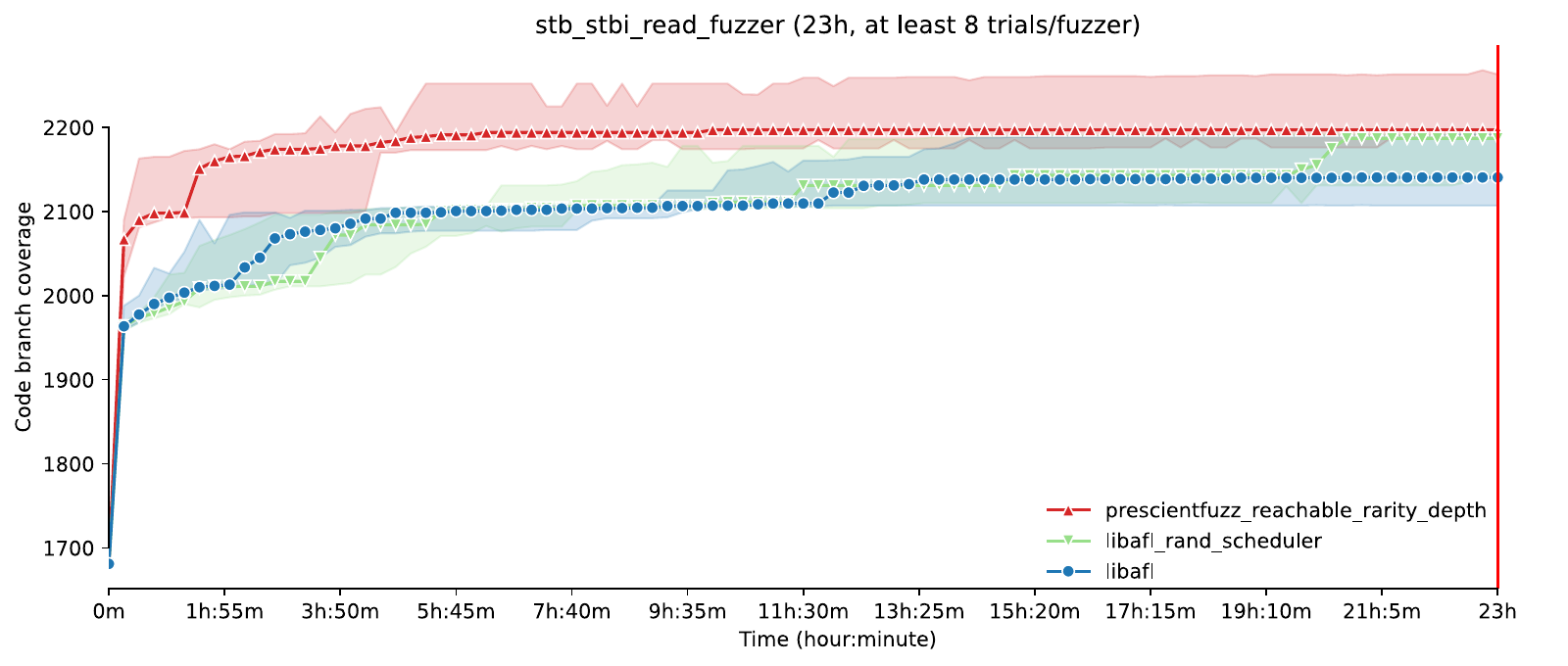}
    \caption{Coverage growth over time for the 2 programs where significant improvement can be seen due to \ourFuzzer's scheduling. The solid line indicates the median coverage, while the shaded area encompasses the 25th and 75th percentiles.}
    \label{fig:did_great}
\end{figure}

There were two programs where a significant improvement could be seen due to the scheduling approach taken by \ourFuzzer; graphs of coverage over time for these are shown in figure \ref{fig:did_great}.

In the case of \texttt{mbed\_fuzz\_dtlsclient}, we see that while `libafl\_rand\_scheduler' makes the initial breakthrough from the plateau around 2,600 edges earlier than `prescientfuzz' (with median at approximately 5.5 hours into the campaign as opposed to 9 hours); the latter continues making more progress towards the end of the time limit.
The AFLFast-based schedule used by `libafl' underperforms here compared to random -- having just tested against this default setup at first, we believed `prescientfuzz' to have made a more significant impact.

In \texttt{stb\_stbi\_read\_fuzzer}, it can be seen that `prescientfuzz' achieves median coverage of approximately 2,070 edges within 15 minutes (points on the plot are at 15 minute intervals); it takes the alternative schedulers over 2 hours to achieve the same coverage.
In this case, the random scheduler catches up towards the end.

The most interesting observation from our point of view is that the random scheduler outperformed the default AFLFast-based setup.
The implementation used in LibAFL's \emph{fuzzbench} fuzzer only mutates a subset of the testcases; these are selected to cover all blocks seen in the campaign while minimising the execution time and input lengths (formerly known as `favored' testcases in AFL).
It is possible that the reduced diversity due to this pruning causes the decrease in rate of coverage discovery.

\begin{figure}[t]
    \begin{minipage}[t]{\linewidth}
        \includegraphics[width=\linewidth]{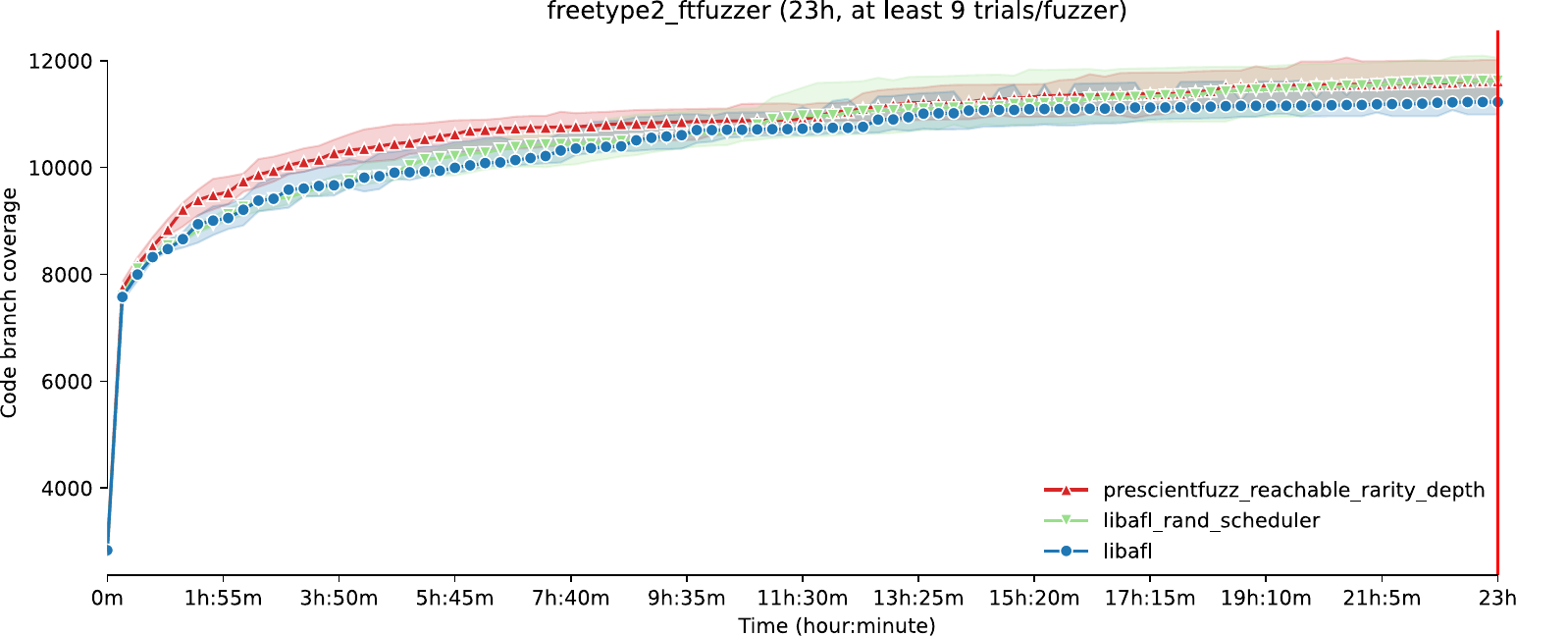}
    \end{minipage}
    \begin{minipage}[t]{\linewidth}
        \includegraphics[width=\linewidth]{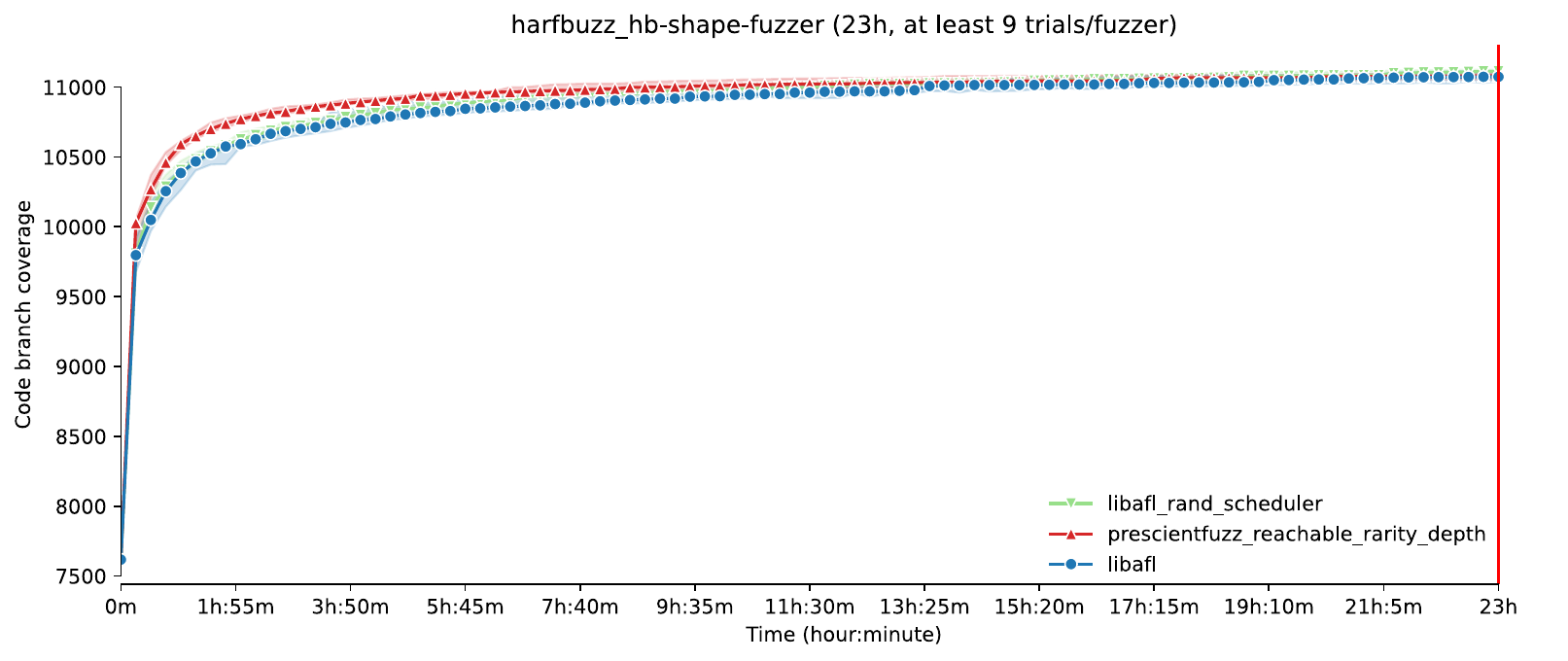}
    \end{minipage}
    \begin{minipage}[t]{\linewidth}
        \includegraphics[width=\linewidth]{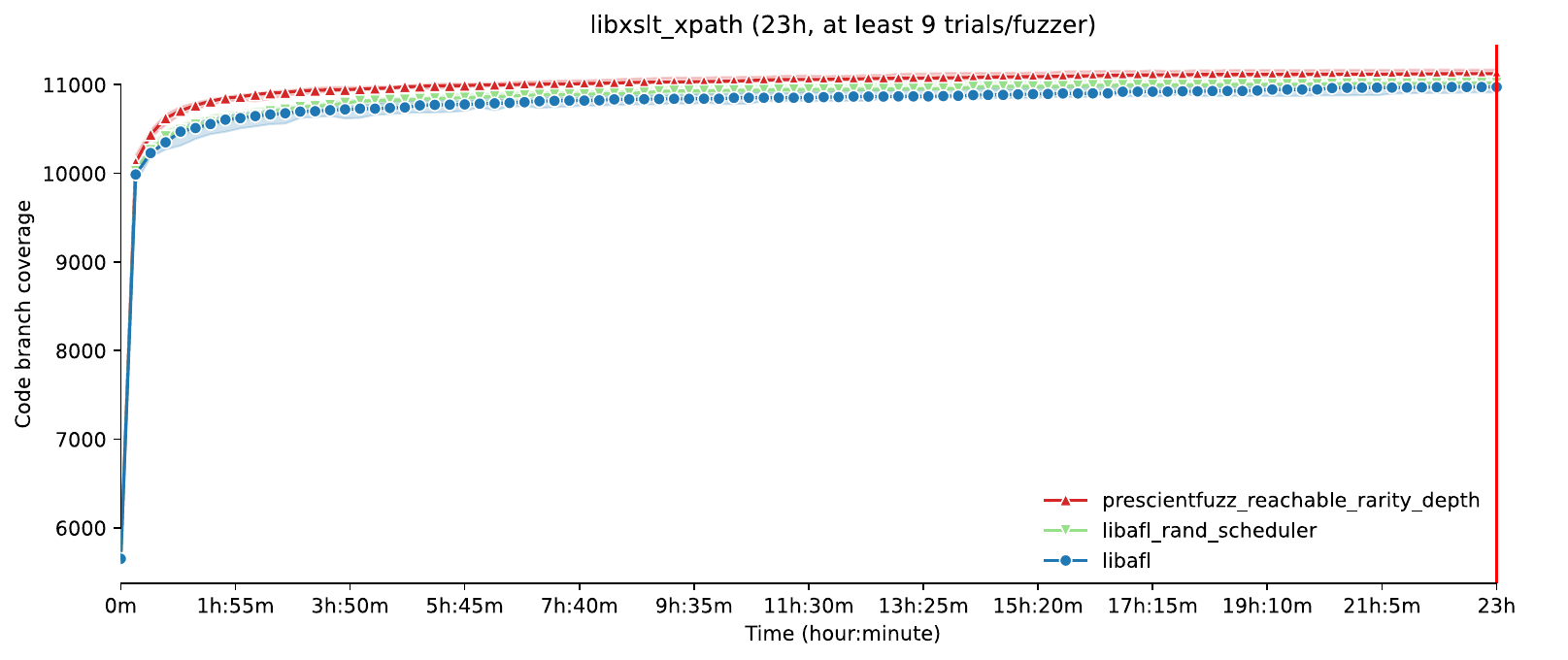}
    \end{minipage}
    \begin{minipage}[t]{\linewidth}
        \includegraphics[width=\linewidth]{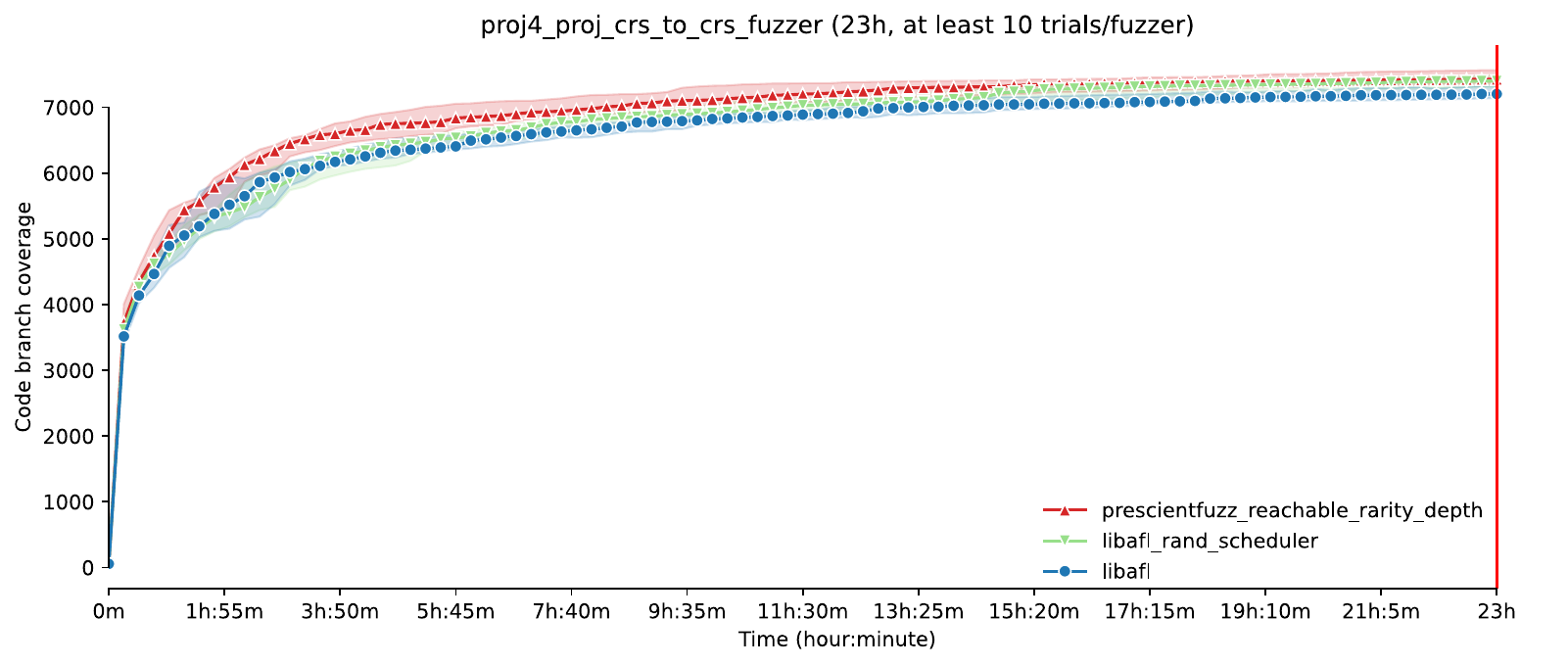}
    \end{minipage}
    
    \caption{Coverage growth over time for four benchmarks where smaller early improvements can be seen due to \ourFuzzer's scheduling; but other scheduling approaches catch up.}
    \label{fig:did_good}
\end{figure}

Figure \ref{fig:did_good} shows four programs where a slight early gain in coverage is seen for \ourFuzzer.
In the case of \texttt{harfbuzz\_hb-shape-fuzzer}, the random scheduler ultimately goes on to slightly outperform \ourFuzzer.
We believe that the flattening of the asymptote, and the number of edges that it occurs at, is due to the power of the mutation engine.
In the case of RedQueen, feedback about the program's internal state -- the value of variables used in conditionals -- essentially acts a powerful hint, `turbocharging' the mutation engine.
As \ourFuzzer{} does not modify the mutation engine nor provide any additional feedback that could guide it, we did not expect an improvement to the number of edges reached when the asymptote flattens -- only an improvement to the time at which the flattening is reached.

\begin{figure}[t]
    \centering
    \includegraphics[width=\linewidth]{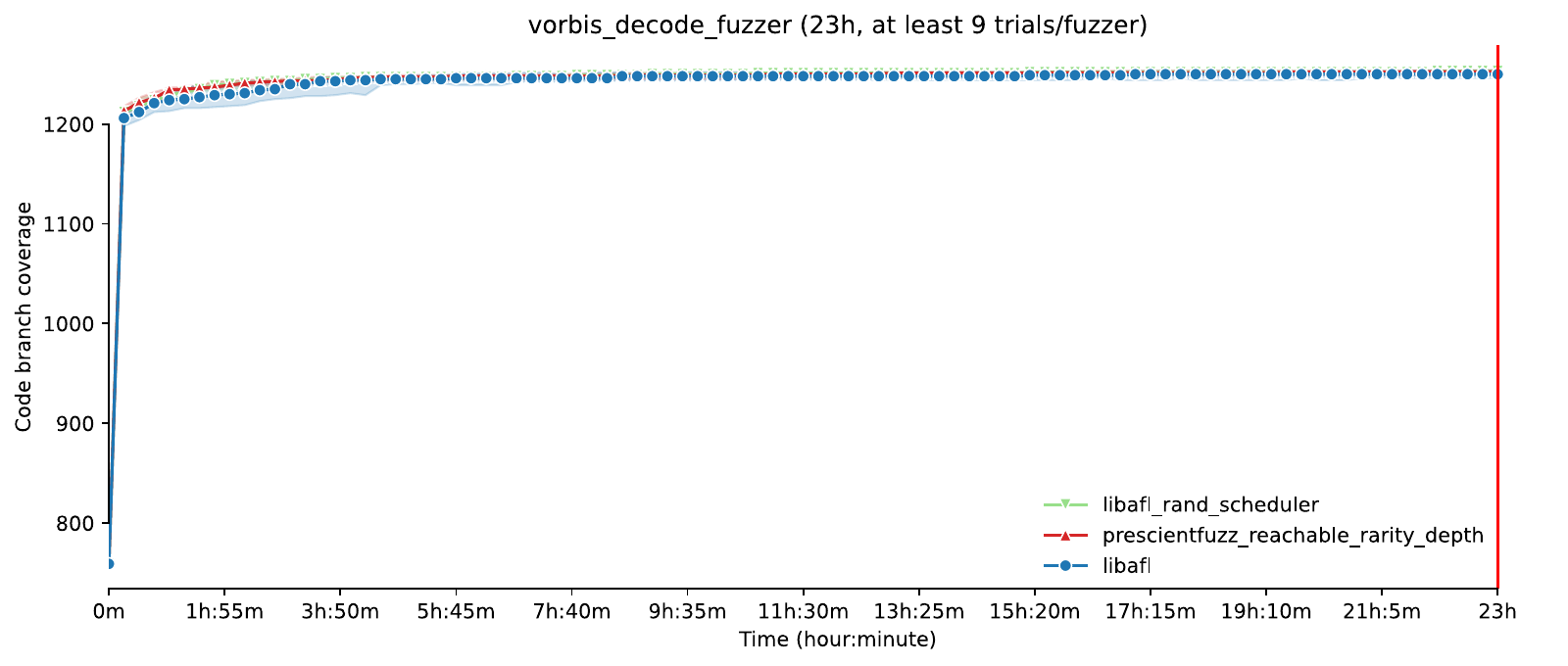}
    \includegraphics[width=\linewidth]{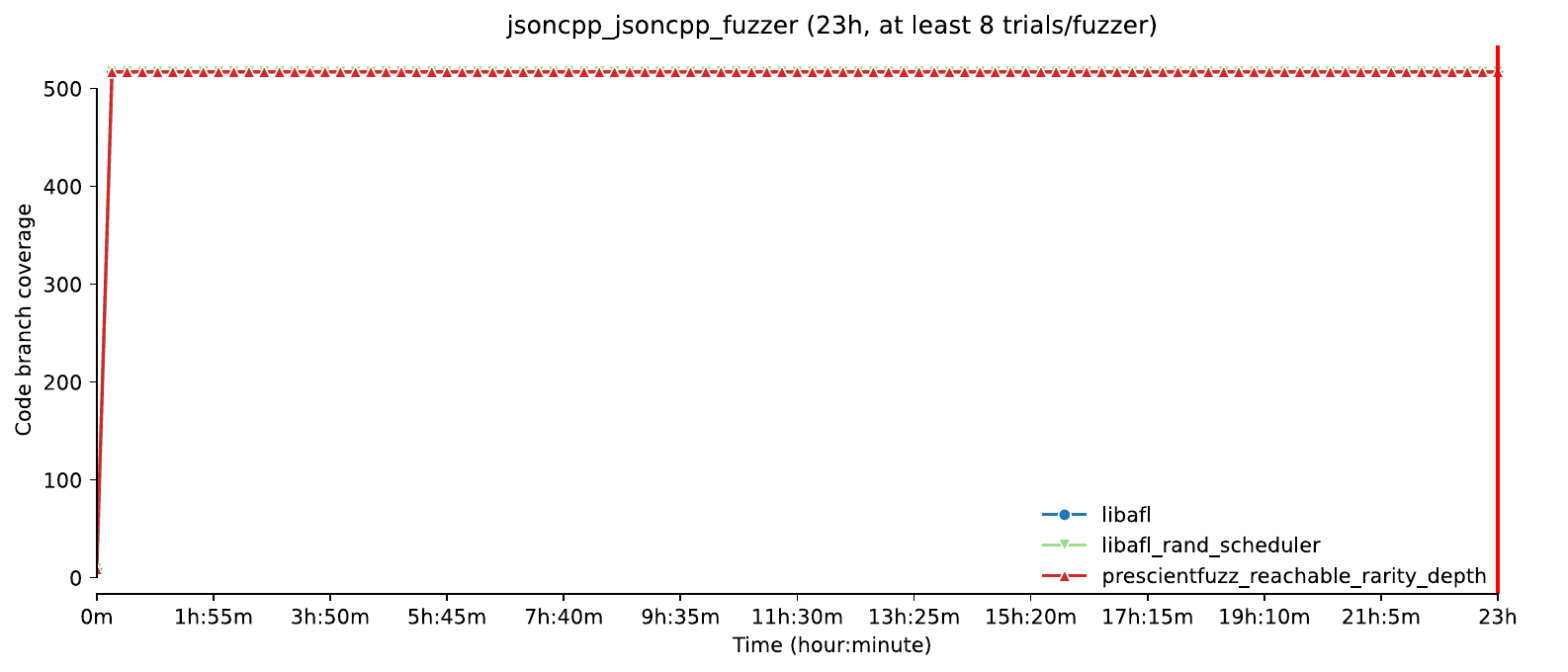}
    \caption{Two benchmarks where there was very little difference observed between schedulers at the 15-minute tick granularity of FuzzBench. Notice that jsoncpp (bottom) is completely flat from 15 minutes onwards.}
    \label{fig:no_winners}
\end{figure}

Finally, Figure \ref{fig:no_winners} shows coverage growth for two SUTs where schedule has little impact.
Notice that in the \texttt{jsoncpp} benchmark, the asymptote has almost completely flattened out by 15 minutes; the time that the first measurement was taken.
We observed that there were no benchmarks where \ourFuzzer{} lagged behind in median coverage at the 23 hour mark by more than 1.0\%.

\section{Impact}

\ourFuzzer{} happens to be the best at discovering coverage on the FuzzBench set of benchmarks, but that is mainly due to `standing on the shoulders of giants'; LibAFL's `fuzzbench' (upon which \ourFuzzer{} is built) was previously the best performing fuzzer.
We see the real value of this work as the \emph{control flow graph} feedback and scheduling mechanism; which can be applied to any grey-box fuzzer.
Given that the compiler pass is already written for LLVM, programs written in C, C++, Rust, and any other languages for which an LLVM front-end exists can benefit from our implementation immediately.

The reachability metric as described in section \ref{sec:reachable}---in particular the set of \emph{All Reachable Uncovered Blocks} (ARUB in Section \ref{sec:formal})---could be used to provide feedback to help determine when to end a fuzzing campaign; improving the estimates provided by some papers \cite{liyanage2024extrapolating,bohme2018stads}.
For example, seeing that coverage discovery has stalled but there still exists a large area of undiscovered coverage (indicated by a large number of reachable blocks) may mean that it is worth continuing to fuzz.
If instead coverage discovery has stalled, but the only remaining uncovered blocks are at depth 1 (i.e. are direct neighbours to covered blocks) then it may be wiser to end the fuzzing campaign.

Additionally, the reachability metric could be useful in determining which branching conditions are best candidates to be solved for when using concolic fuzzing.
Concolic fuzzing uses a concrete input to build up a set of constraints that are required to reach a certain program point; essentially a formula describing the necessary features of the input required.
We can foresee that sorting the set of basic blocks for each input by the number of reachable blocks -- Then creating and solving the formula for the block with most reachability -- may vastly improve the efficiency, and get the most value out of each expensive solve.

\section{Threats to Validity}

Outsourcing our evaluation to the hosted FuzzBench service means that we had no control over the execution conditions; despite this, FuzzBench is commonly trusted for the evaluation of fuzzers in academic work \cite{fioraldi2022libafl,bohme2020boosting,mantovani2022fuzzing}.

\section{Conclusion}

In this chapter we have introduced a new dynamic feedback mechanism for grey-box (and white-box) fuzzers, that makes use of the SUT's CFG semantics.
We also document the design of an input corpus scheduler, that combines this knowledge from the CFG with coverage feedback from the corpus itself, in order to select inputs for mutation that have a higher probability of discovering new coverage.
We built a grey-box fuzzer, \ourFuzzer, and evaluated it on FuzzBench, where it achieves the most coverage across the aggregated set of benchmarks of any publicly available fuzzer.
Finally, we discussed how the feedback mechanism could be used to improve decisions about when a fuzzing campaign should be halted, and how it could be used by concolic fuzzers in order to improve the leverage of the computationally expensive solving process by intelligently selecting which conditionals to solve.


\bibliographystyle{splncs04}
\bibliography{references}

\end{document}